\documentclass[]{aastex631}

\begin{document}

\title[QPOs of GX 339$-$4]{Quasi-periodic oscillations in GX 339$-$4 during the 2021 Outburst observed with \textit{Insight-HXMT}}

\author{Y. J. Jin}
\affiliation{Department of Astronomy, School of Physics and Technology, Wuhan University, Wuhan 430072, China}
\affiliation{WHU-NAOC Joint Center for Astronomy, Wuhan University, Wuhan 430072, China}

\author{W. Wang}
\altaffiliation{Email address: wangwei2017@whu.edu.cn}
\affiliation{Department of Astronomy, School of Physics and Technology, Wuhan University, Wuhan 430072, China}
\affiliation{WHU-NAOC Joint Center for Astronomy, Wuhan University, Wuhan 430072, China}

\author{X. Chen}
\altaffiliation{Email address: xiao.chen@whu.edu.cn}
\affiliation{Department of Astronomy, School of Physics and Technology, Wuhan University, Wuhan 430072, China}
\affiliation{WHU-NAOC Joint Center for Astronomy, Wuhan University, Wuhan 430072, China}

\author{P. F. Tian}
\affiliation{Department of Astronomy, School of Physics and Technology, Wuhan University, Wuhan 430072, China}
\affiliation{WHU-NAOC Joint Center for Astronomy, Wuhan University, Wuhan 430072, China}
\author{Q. Liu}
\affiliation{Department of Astronomy, School of Physics and Technology, Wuhan University, Wuhan 430072, China}
\affiliation{WHU-NAOC Joint Center for Astronomy, Wuhan University, Wuhan 430072, China}
\author{P. Zhang}
\affiliation{Department of Astronomy, School of Physics and Technology, Wuhan University, Wuhan 430072, China}
\affiliation{WHU-NAOC Joint Center for Astronomy, Wuhan University, Wuhan 430072, China}
\author{H. J. Wu}
\affiliation{Department of Astronomy, School of Physics and Technology, Wuhan University, Wuhan 430072, China}
\affiliation{WHU-NAOC Joint Center for Astronomy, Wuhan University, Wuhan 430072, China}
\author{N. Sai}
\affiliation{Department of Astronomy, School of Physics and Technology, Wuhan University, Wuhan 430072, China}
\affiliation{WHU-NAOC Joint Center for Astronomy, Wuhan University, Wuhan 430072, China}



\begin{abstract}

A new outburst of GX 339$-$4 in 2021 was monitored by the Hard X-ray Modulation Telescope (\textit{Insight}-HXMT). By using the data of \textit{Insight}-HXMT from February to March 2021, we make the X-ray timing analysis of this new outburst. Based on the results of count rates, hardness-intensity diagram (HID) and power density spectrum (PDS), we confirm that the source exhibits spectral transitions from the low-hard state (LHS) to the hard-intermediate state (HIMS). During the transition from the LHS to the HIMS, Low-frequency Quasi-periodic oscillations (LFQPOs) are detected in the PDS. We found that these QPOs are all type-C QPOs with centroid frequencies evolving from $0.1 -0.6$ Hz in the LHS and in the $1 -3$ Hz frequency range in HIMS. The QPO features above 50 keV are reported for the first time in this black hole by \textit{Insight}-HXMT. The QPO rms stays stable with time but decreases with energy at higher energy above $\sim 10$ keV. We also find that the phase lag of the type-C QPO is close to zero in the early outburst stage, but becomes positive as the outburst evolves, with a hard lag of $\sim$ 0.6-1.2 rad in $50 -100 $ keV. The implications of the phase lag in high energy bands and possible physical mechanisms to explain those observations are also discussed.

\end{abstract}

\keywords{accretion: accretion disks --- black hole physics --- X-rays: binaries --- stars: individual: GX 339$-$4 }


\section{Introduction} 
\label{sec:in}
Low mass X-ray binaries (LMXRBs) consist of a compact object (black hole or neutron star) and a low-mass companion star. The systems are powered by the accreting material from the companion star through Roche lobe overflow and release most of their energy in X-rays. Black hole XRBs (BHXRBs) are generally transient X-ray systems that spend most of their lives in quiescence and sometimes go into an outburst that can last for a few months \citep{Remillard2006}. During the outbursts, BHXRBs normally follow the spectral transition \citep{Belloni2005} from the low-hard state (the LHS), the hard-intermediate state (the HIMS), the soft-intermediate state (the SIMS), to the high-soft state (the HSS), which can form a 'Q' feature in the hardness-intensity diagram (HID).

Low-frequency quasi-periodic oscillations (LFQPOs) with the frequency ranging from a few mHz to $\sim$30 Hz are an important timing feature of BHXRBs \citep{Casella2004}. The study of LFQPOs is essential to understand the accretion flow around black holes, though the origin of LFQPOs is still under debate \citep{Ingram2009}. Generally, the quasi-periodic oscillations (QPOs) observed in BHXRBs are classified into Type-A, -B, -C QPOs based on properties like the quality factor (Q=$\nu_{\rm centroid}/{\rm FWHM}$) and the shape of the noise associated with the oscillation \citep{Motta2011}.
Type-C QPOs are the most common type of QPOs in BHXRBs, and occur mostly in the LHS and the HIMS \citep{Casella2005}. The QPOs are characterized by a strong, narrow peak with variable frequency, superposed on a Flat-Top Noise (FTN) that steepens above a frequency comparable to the QPO frequency. A second harmonic and a subharmonic peak are sometimes present in the power density spectrum (the PDS, \citealt{Zhang2017}).

One of the promising scenarios for LFQPOs (in particular type-C QPOs) postulates that the oscillations arise from the Lense-Thirring precession \citep{Stella1998} of the misaligned inner hot flow\citep{Ingram2009,Ingram2010,Ingram2011}. In this geometry assumption, the inner accretion flow forms a hot, geometrically thick, optically thin configuration, while the outer accretion flow takes the form of a cool, geometrically thin, optically thick accretion disk truncated at some large radius \citep{Motta2011}.
Evidence in support of this scenario has been inferred from some timing features in BHXRBs, e.g., the modulation of the reflected iron line equivalent width\citep{Ingram2015}, the centroid energy \citep{Ingram2016} during a QPO cycle using phase-resolved spectroscopy of type-C QPOs, the inclination dependence of QPO phase lags \citep{Eijnden2017}, and the absolute variability amplitude \citep{Motta2015}. However, as indicated by \cite{Marcel2021}, this model may not be consistent with the realistic accretion flow properties at all times. A disk-jet scenario under the framework of hybrid disk radial zones is thus proposed to explain the existence of QPOs in X-rays, UV and IR, and successfully matches most of the observed data for GX 339$-$4 \citep{Ferreira2022A&A}.

The energy dependence of the QPO properties is also important for QPO origin study, such as fractional rms, centroid frequency, and time-lag \citep{Tomsick2001,Rodriguez2004}. Based on the inclination dependence of phase lags in a sample of 15 BHXRBs, \citet{Eijnden2017} find that the phase lags of type-C QPOs strongly depend on inclination, and such an inclination dependence provides strong constraints on the physical mechanism of QPOs. Recently, a new Comptonization model \citep{Karpouzas2020} was proposed to study the energy dependence of fractional rms and phase lags for both type-B \citep{García2021} and type-C \citep{Karpouzas2021,zhang2022MNRAS,Rawat2023} QPOs.

GX 339$-$4 is a recurrent low-mass X-ray binary discovered in 1973 \citep{Markert1973}, which harbors a black hole with a mass function of f(M) = 5.8 $\pm 0.5 M_{\sun}$ \citep{Hynes2003}. The distance to this system is between 6 and 15 kpc \citep{Hynes2004}. In addition, GX 339$-$4 is seen at relatively low binary and inner disk inclinations ($i \leq 45^\circ$, \citealt{Nowak2002,Hynes2003}). GX 339$-$4 has undergone frequent outbursts and displayed all the black hole accretion states \citep[e.g.][]{Belloni1999,Belloni2005,Dunn2008,Zhang2017} in the past thirty years, and has been extensively studied at all wavelengths, making it one of the most studied BHXRBs. Strong QPOs were often observed in the PDS of GX 339$-$4 during its outbursts.
\citet{Miyamoto1991} reported the QPO centroid frequency evolving from $1-15$ Hz together with a strong variable band-limited noise observed with Ginga.
\citet{Nespoli2003} reported a transient 6 Hz QPO from GX 339$-$4 with the Rossi X-ray Timing Explorer (RXTE) during the 2002/2003 outburst.
\citet{Motta2011} studied the properties and behaviors of LFQPOs during four outbursts (in 2002, 2004, 2007, 2010). Most of the QPOs could be classified as the types of ABC that have been proposed before. During the 2010 outburst, \citet{Kalamkar2016} reported the detection of a QPO at 0.08 Hz in the infra-red (IR) band, which is half of the X-ray QPO frequency at 0.16 Hz; and a weak sub-harmonic close to the IR QPO frequency was also detected in X-rays.

GX 339$-$4 entered a new outburst in 2021, which was observed by \textit{Insight}-HXMT. The HID of this outburst \citep{Liu2021} showed that the source spent some time in the hard state followed by a fast transition (around a week) to the soft state. At the end of this transition, this source showed rapid changes of flux in the hard X-ray band. This variability happened on a timescale of less than one orbit of \textit{Insight}-HXMT (1.5 hours). After that, the source entered the very high state and then the soft state. Tracing the evolution of rapid X-ray variability along an outburst can help us understand the physical changes of the accretion flow and the origin of the variability.

In this paper, we study in details the evolution of the type-C QPOs in GX 339$-$4 during the low hard state of the 2021 outburst based on Insight-HXMT observations. In Section~\ref{sec:ob}, we describe the observations and data reduction methods. The QPO evolution and properties are presented in Section~\ref{sec:re}. Our discussion and main conclusions follow in Sections~\ref{sec:di} and~\ref{sec:co}, respectively.

\section{Observation and data analysis} 
\label{sec:ob}
\textit {Insight}-HXMT is the first X-ray satellite of China, with three detectors onboard, i.e. the High-Energy X-ray telescope (HE), the Medium-Energy X-ray telescope (ME), and the Low Energy X-ray telescope (LE), covering a very wide range of energy band from 1 keV to 250 keV \citep{Zhang2020}. HE has 18 NaI/CsI detectors covering the energy band of 20-250 keV, ME has 1728 Si-PIN detectors covering 5-30 keV, while LE uses Swept Charge Device to detect the energy range of 1-15 keV. The typical fields of view are: 1.1$^\circ$ × 5.7$^\circ$ for HE, 1$^\circ$ × 4$^\circ$ for ME and 1.6$^\circ$ × 6$^\circ$ for LE \citep{Huang2018}.

\textit{Insight}-HXMT started monitoring the new outburst of GX 339$-$4 from February 18, 2021. We use the Insight-HXMT Data Analysis software (HXMTDAS) v2.04 to analyze all the data. Here some details about data selections are briefly introduced: the pointing offset angle $< 0.04^\circ$, the elevation angle $> 10^\circ$ and geomagnetic cutoff rigidity $> 8^\circ$; data within 300 s of the South Atlantic Anomaly (SAA) passage are not used. We set 0.0078125s time bins for the light curves made by HXMTDAS tasks \textit{helcgen}, \textit{melcgen} and \textit{lelcgen}. The backgrounds of the spectra and light curves are estimated by the official tools (\textit{HEBKGMAP}, \textit{MEBKGMAP} and \textit{LEBKGMAP}) of version 2.0.12. The details on the data analysis and processes can be found in previous works \citep[e.g.][]{WANG2021,Chenxiao2021,Chenxiao2022}.

Then the PDSs are obtained with 128s data intervals and 1/128s time resolution. The PDS is produced with Possion noise subtracted and fractional rms normalization \citep{Belloni1990,Miyamoto1991}, and fitted with several Lorentzian functions using XSPEC v12.12.1. We then obtain the QPO centroid frequency and the fractional rms amplitude so that the evolution of frequency and amplitude with time can be determined.
To show more accurate and representative results, we calculate the significance of QPOs, which is defined as the integral of the LFQPO corresponded Lorentzian divided by its error. Then four representative observations (P0304024007, P0304093012, P0304093024, P0304024036) are selected based on the following criteria for further analysis: (a) large overlapping GTIs for all the three telescopes ($> 1000$ s); (b) high QPO significance ($> 4\sigma$); (c) large time interval between any chosen observations ($>$ one day). These four observations are marked in Figure~\ref{fig:rate} and Figure~\ref{fig:HID} with red stars.

To study the energy dependence of QPOs, we make the PDS in six energy bands (2-4 keV, 4-8 keV, 8-14 keV, 14-30 keV, 30-50 keV and 50-100 keV) and calculate the properties of the QPO in each PDS, so that we can find out the evolution of frequency and amplitude with different energy bands. In addition, we produce the cross spectrum between a reference band (2-4 keV) and several harder bands (4-8 keV, 8-14 keV, 14-30 keV, 30-50 keV and 50-100 keV). The frequency-lag spectra are obtained to study the lag dependence around the QPO frequency. We find out that the PDS are normally dominated by the broad-band noise at the QPO frequency. In order to disentangle the lag of the QPO from that of the broad-band noise, we calculate the QPO lag using the method proposed by \cite{Alabarta2022}, which is the arctan of the ratio of the normalization of the Real and Imaginary parts of the QPO, where the normalizations are obtained by fitting the Real and Imaginary parts of the cross spectrum with the same Lorentzian model used to fit the corresponding PDS, respectively, except the centroid frequencies and FWHM are fixed. The errors of the Real and Imaginary parts are calculated following the formula (13) in \cite{Ingarm2019mn}. The positive lag values mean that the hard photons are lagging the soft ones and negative lag values mean opposite \citep{Zhang2017,Eijnden2017}.

\section{Results} 
\label{sec:re}
\subsection{Hardness-Intensity diagram \label{sec:maths}}
\begin{figure}
\plotone{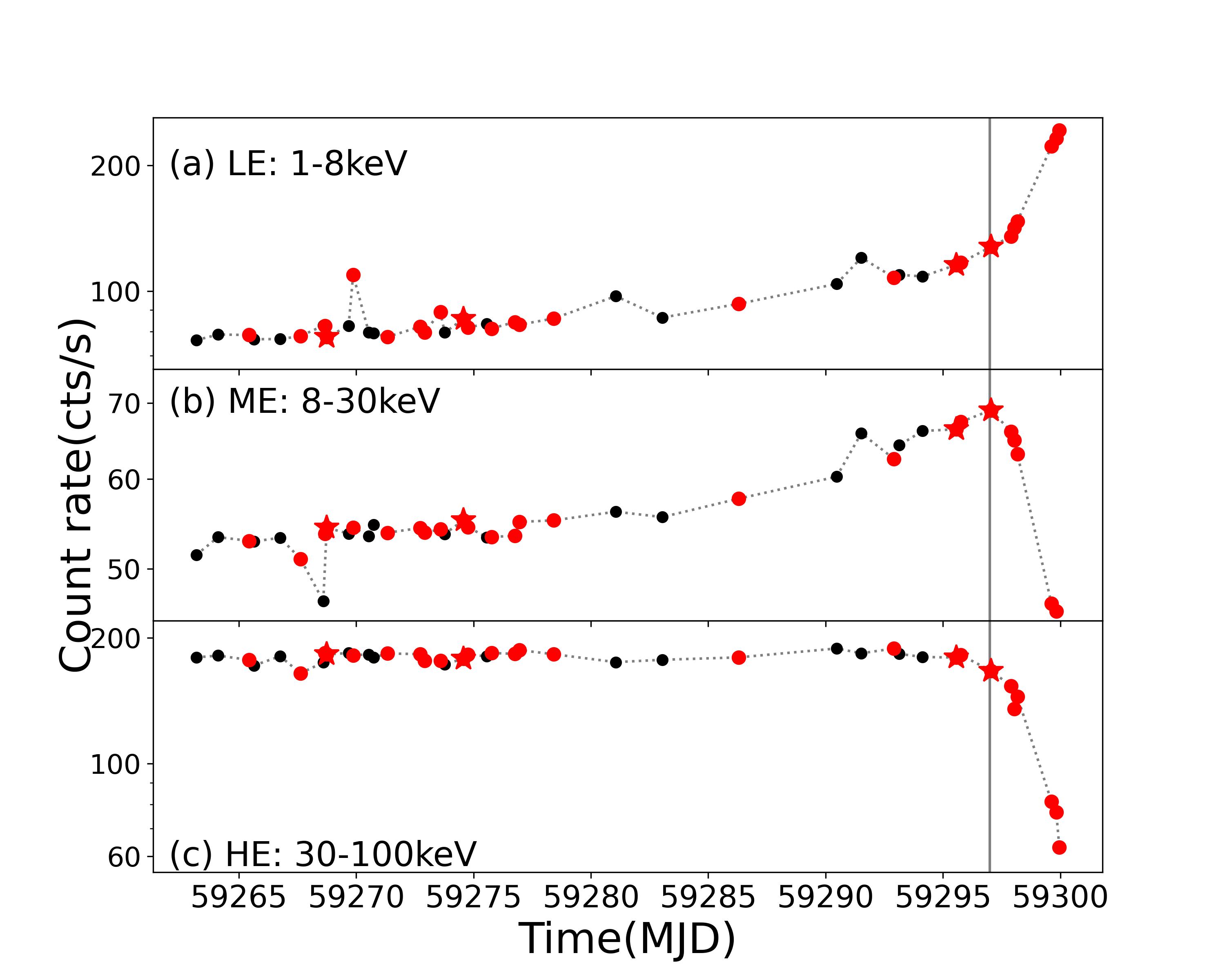}
\caption{LE (1-8 keV), ME (8-30 keV), and HE (30-100 keV) count rate curves of Insight-HXMT. Each point represents one Insight-HXMT observation and the red points represent observations with QPO detected. The four red stars represent the observations analyzed in detail (e.g the energy dependences of the PDS, rms and phase lag) as mentioned in Section 2. The grey line marks the transition from the LHS to the HIMS.
\label{fig:rate}}
\end{figure}

The light curves for 1-8 keV, 8-30 keV and 30-100 keV show the evolution of count rates as a function of time for LE, ME, HE, respectively, in Figure~\ref{fig:rate}. The LE count rate (1-8 keV) rises slowly from its initial value to MJD 59297, and then increases rapidly. The ME count rate (8-14 keV) stays stable at $\sim 53$ cts $s^{-1}$ from MJD 59263 to 59276, and slowly increases to 69 cts $s^{-1}$ on MJD 59297, then decreases abruptly to 46 cts $s^{-1}$ on MJD 59299. The HE count rate (30-100 keV) shows a constant rate of $\sim 180$ cts $s^{-1}$ in the early phase and an abrupt decrease also happens on MJD 59297.
It is clear that the evolution of count rates can be roughly divided into two stages. The count rates increase slowly in the first stage (MJD 59263-59297), which is considered to be in the LHS. While in the second stage (after MJD 59297), the count rates begin to change abruptly, increasing in LE but decreasing in ME and HE, which represents the beginning of the state transition from the LHS to the HIMS.

\begin{figure}
\plotone{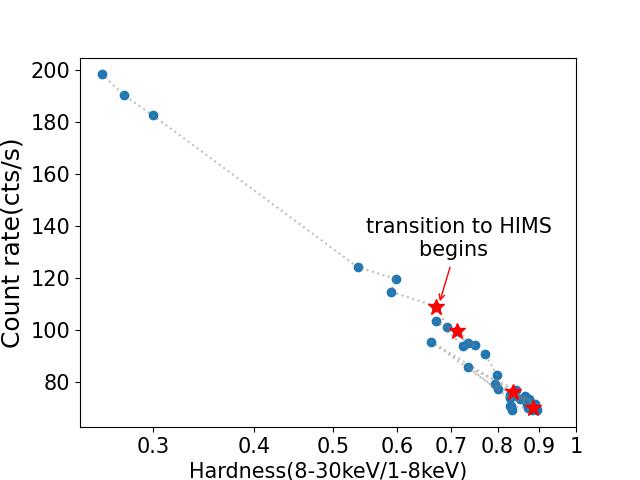}
\caption{The hardness intensity diagram of GX 339$-$4 during the 2021 outburst. The Y-axis represents the count rate of LE from $1-8$ keV. The hardness ratio is defined as the count ratio between the energy bands $8-30$ keV and $1-8$ keV. Each point represents one Insight-HXMT observation. The four red stars represent the observations analyzed in detail (e.g the energy dependences of the PDS, the QPO rms and phase lag) as mentioned in Section 2.
\label{fig:HID}}
\end{figure}

Hardness is also calculated to make the hardness-intensity diagram (HID, see Figure~\ref{fig:HID}). Hardness is defined as the count rate ratio of two energy bands: $8-30$ keV over $1-8$ keV, and intensity is defined as the count rate of LE from $1-8$ keV. The complete loop in HID usually presents a Q-shaped form. In this work, the observation data are from the early stage of the 2021 outburst, including the state evolution from LHS to HIMS. The outburst starts at the lower right corner of the diagram. Before MJD 59297, hardness and intensity are changing in a small range, causing the observations to be concentrated in the lower right corner of the figure. After that, the outburst moves to the upper left with decreasing hardness and increasing intensity. A relatively complete evolution pattern of GX 339$-$4 during its 2021 outburst is reported by \citet{Liu2021} with \textit{Insight}-HXMT data.

\subsection{QPO properties}
\begin{figure}
\plotone{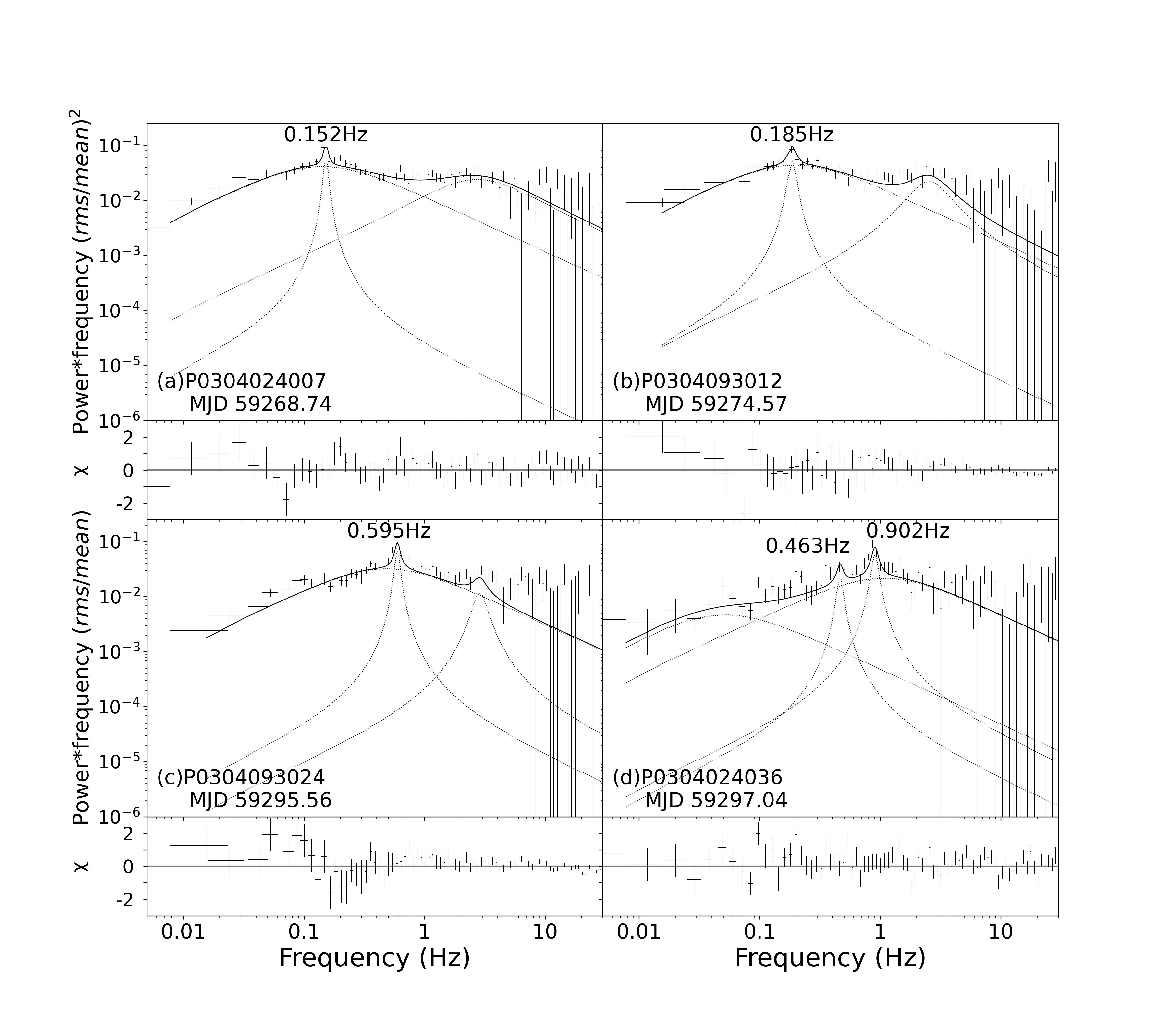}
\caption{The power density spectra for the four representative observations using the Insight-HXMT/LE data (1-8 keV). The solid lines show the best fit with a multi-Lorentzian function (dotted lines). Observation ID and QPO fundamental frequency are shown for each panel. Panel (a), (b) and (c) show the PDS in the LHS, while the representative PDS in the transition from the LHS to the HIMS is shown in panel (d).}
\label{fig:pds-time}
\end{figure}

For each observation, we obtain the PDS and then fit them with multiple Lorentzian profiles as mentioned in Section~\ref{sec:ob}. In Figure~\ref{fig:pds-time}, we show the PDS of four representative observations as mentioned in Section 2, using their LE data (1--8 keV). The PDS normally can be fitted with one narrow Lorentzian component which is the typical type-C QPO, and two broad Lorentzian components. As the source evolved from the LHS to the HIMS, the QPO is constrained better and the centroid frequency is higher. Furthermore, sub-harmonics can be noticed in the PDS during the state transition (see panel (d) in Figure~\ref{fig:pds-time}). Similar QPO properties have also been found in other black holes during their LHS at the early stage of the outburst \citep{Belloni2005,Huang2018}. In Table~\ref{tab:table1}, we present a summary of the observations with significant ($>3\sigma$) QPO signals detected, together with the fitting parameters of QPOs (i.e., the centroid frequency $\upsilon_{\rm QPO}$, the FWHM, the rms) with 1 $\sigma$ error using the LE data of 1-8 keV.

\begin{table}
    \centering
    \caption{Low-frequency QPO parameters for GX 339$-$4 using the LE (1-8 keV) data of \textit{Insight}-HXMT. Only observations with QPO significance $ >3\sigma$ are shown.}
    \label{tab:table1}
    \begin{tabular}{ccccccccc}
    \hline
    \hline
        &&&&&&&&Harmonic\\
        &&&&Fundamental&&&&or\\
        &&&&&&&&Sub-harmonic\\
    \hline
        No. &Observation ID & MJD & Exposure & $\nu_{\rm QPO}$ & FWHM & rms & significance & $\nu_{\rm QPO}$ \\
        &&&(s)&(Hz)&(Hz)&(\%)& &(Hz)\\
    \hline
        \\1 & P0304093001 & 59265.43 & 831 & $0.150_{-0.005}^{+0.007}$ & $0.035_{-0.015}^{+0.014}$ & $15.6_{-2.4}^{+2.1}$ & 3.45 &-\\
        \\2 & P0304024007 & 59268.74 & 4055 & $0.152_{-0.002}^{+0.002}$ & $0.013_{-0.005}^{+0.006}$ & $9.3_{-1.3}^{+1.0}$ & 4.22 &-\\
        \\3 & P0304093011 & 59273.59 & 3995 & $0.178_{-0.003}^{+0.001}$ & $0.010_{-0.005}^{+0.007}$ & $6.3_{-0.6}^{+1.0}$ & 3.73 &-\\
        \\4 & P0304093012 & 59274.57 & 2700 & $0.185_{-0.001}^{+0.001}$ & $0.028_{-0.010}^{+0.015}$ & $11.5_{-1.5}^{+1.4}$ & 4.04 &-\\
        \\5 & P0304024016 & 59275.76 & 2867 & $0.180_{-0.001}^{+0.001}$ & $0.012_{-0.007}^{+0.005}$ & $7.7_{-1.4}^{+1.1}$ & 3.13 &-\\
        \\6 & P0304093014 & 59276.75 & 3202 & $0.187_{-0.001}^{+0.001}$ & $0.014_{-0.007}^{+0.017}$ & $8.2_{-1.3}^{+1.2}$ & 3.39 &-\\
        \\7 & P0304024017 & 59276.95 & 1648 & $0.185_{-0.001}^{+0.001}$ & $0.018_{-0.010}^{+0.016}$ & $8.1_{-1.9}^{+0.9}$ & 3.09 &-\\
        \\8 & P0304093016 & 59278.41 & 2390 & $0.190_{-0.001}^{+0.001}$ & $0.016_{-0.006}^{+0.010}$ & $8.9_{-1.5}^{+1.1}$ & 3.45 &-\\
        \\9 & P0304093021 & 59286.29 & 720 & $0.287_{-0.006}^{+0.005}$ & $0.055_{-0.012}^{+0.017}$ & $12.4_{-1.3}^{+1.2}$ & 5.12 &-\\
        \\10 & P0304093023 & 59292.91 & 1380 & $0.512_{-0.004}^{+0.006}$ & $0.029_{-0.014}^{+0.019}$ & $7.2_{-1.1}^{+1.1}$ & 3.29 &-\\
        \\11 & P0304093024 & 59295.56 & 1967 & $0.595_{-0.005}^{+0.007}$ & $0.068_{-0.017}^{+0.028}$ & $10.9_{-0.5}^{+1.0}$ & 6.95 &-\\
        \\12 & P0304024035 & 59295.76 & 2086 & $0.655_{-0.011}^{+0.012}$ & $0.121_{-0.029}^{+0.035}$ & $10.8_{-1.2}^{+1.0}$ & 4.87 &-\\
    \hline
        \\13 & P0304024036 & 59297.04 & 1200 & $0.902_{-0.013}^{+0.012}$ & $0.134_{-0.037}^{+0.030}$ & $11.4_{-1.2}^{+1.0}$ & 5.19 & 0.463\\
        \\14 & P0304024037-1 &59297.91 & 840 & $1.486_{-0.008}^{+0.008}$ & $0.071_{-0.021}^{+0.027}$ & $8.8_{-0.8}^{+0.8}$ & 5.28  & -\\
        \\15 & P0304024037-2 &59298.04 & 900 & $1.683_{-0.025}^{+0.022}$ & $0.166_{-0.053}^{+0.064}$ & $8.1_{-0.9}^{+1.0}$ & 4.24 & 0.891\\
        \\16 & P0304024037-3 &59298.17 & 960 & $1.959_{-0.030}^{+0.030}$ & $0.260_{-0.067}^{+0.073}$ & $9.0_{-1.2}^{+1.0}$ & 4.17 & 0.99\\
        \\17 & P0304024038-1 &59299.62 & 2633 & $2.555_{-0.027}^{+0.027}$ & $0.387_{-0.081}^{+0.084}$ & $7.8_{-0.7}^{+0.4}$ & 7.67 & 5.44\\
    \hline
    \end{tabular}
\end{table}

\begin{figure}
\plotone{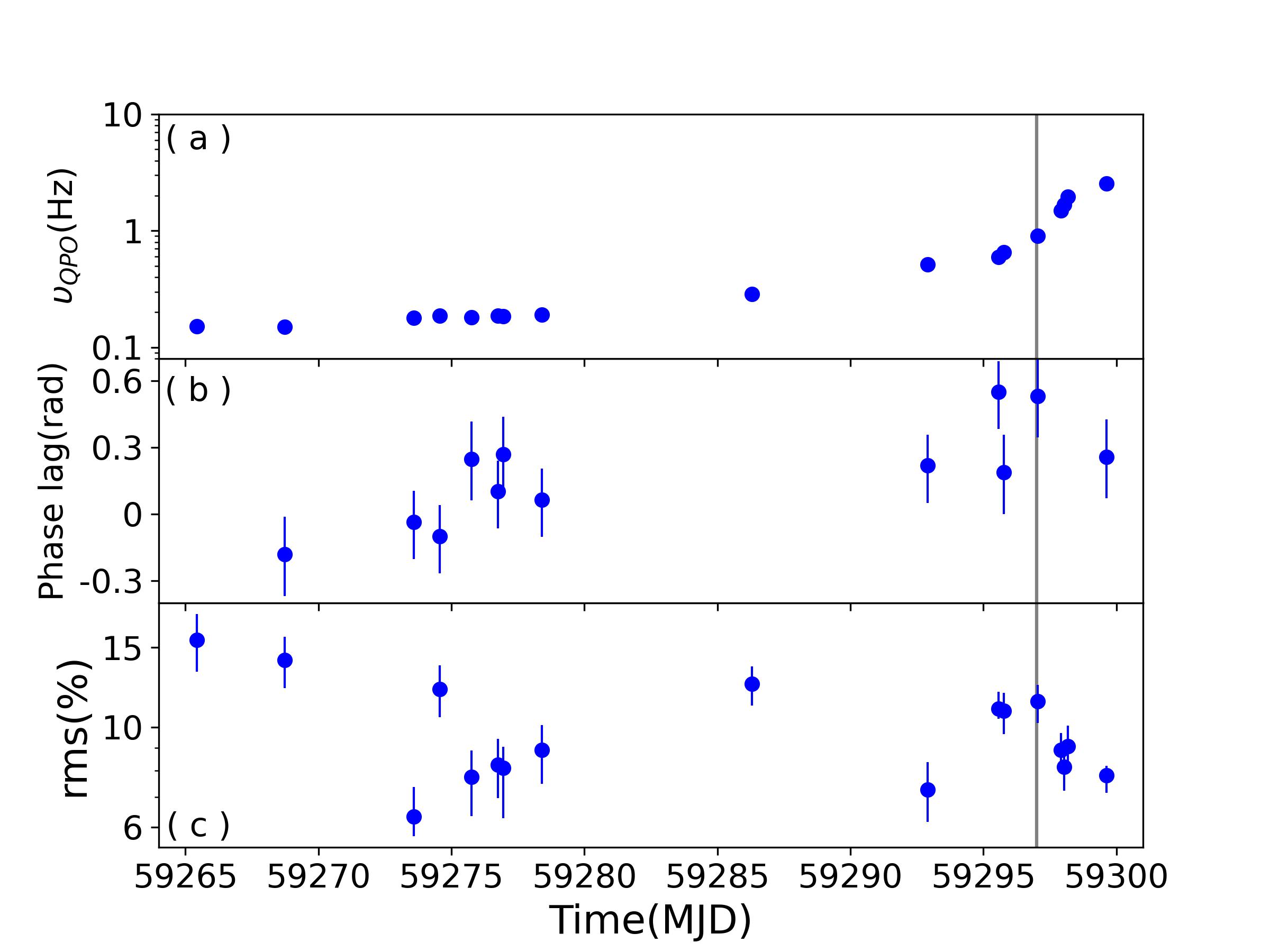}
\caption{Evolution of the QPO frequencies (a) and the QPO rms (c) of GX 339$-$4 in the band of $1-8$ keV as a function of time. Phase lags (b) are calculated between 2 - 4 keV and 30 - 50 keV. Only the phase lag for observations with overlapping exposure greater than 1000 s are listed. The solid line marks the transition from the LHS to the HIMS.}
\label{fig:time}
\end{figure}

Figure~\ref{fig:time} shows the evolution of (a) the QPO centroid frequency, (b) phase lag at the QPO frequency and (c) the QPO rms, as a function of time, where phase lag is measured between $2-4$ keV and $30-50$ keV. In Figure~\ref{fig:time}(a), the centroid frequency of the QPO increases from 0.1 Hz to 3 Hz in our sample. Similar to the count rate evolution of LE, the centroid frequency of QPO increases slowly before MJD 59297, but increases rapidly from 0.9 Hz to 3 Hz within three days when the transition from the LHS to the HIMS begins. A similar frequency evolution of GX 339$-$4 occurred during its 2006/2007 outburst as reported by \citet{Zhang2017}.
In Figure~\ref{fig:time}(c), the QPO rms has no obvious trend with time. Due to the relatively poor constraints on QPOs in the early stage of outburst, the rms values are widely distributed in the range of $\sim 4-15\%$ with large errors. As the QPOs are constrained better, the rms values are constrained around 10\%.

The phase lag evolving with time is shown in Figure~\ref{fig:time}(b) and Table~\ref{tab:table3}. Only the phase lags for observations with overlapping exposure more than 1000 s are shown in this paper. A gradual increase of the phase lag over time can be noticed. 

\begin{table}
    \centering
    \caption{Phase lags at the QPO fundamental for GX 339$-$4. Lags are calculated between 2-4 keV and 30-50 keV. The exposure time is the length of the overlapping GTIs of the two energy bands. Only observations with overlapping exposure greater than 1000 s are listed.}
    \label{tab:table3}

    \begin{tabular}{cccccc}
    \hline
    \hline
        No. &Observation ID & MJD & Exposure & $\nu_{\rm QPO}$ & Phase lag \\
        &&&(s)&(Hz)&(rad)\\
    \hline
        \\1& P0304024007& 59268.74& 4055&$0.152_{-0.002}^{+0.002}$& $-0.182_{-0.123}^{+0.129}$\\
        \\2& P0304093011& 59273.59& 3995&$0.178_{-0.003}^{+0.001}$& $-0.035_{-0.207}^{+0.210}$\\
        \\3& P0304093012& 59274.57& 2700&$0.185_{-0.001}^{+0.001}$& $-0.099_{-0.131}^{+0.135}$\\
        \\4& P0304024016& 59275.76& 1967&$0.180_{-0.001}^{+0.001}$& $0.247_{-0.192}^{+0.175}$\\
        \\5& P0304093014& 59276.75& 3202&$0.187_{-0.001}^{+0.001}$& $0.102_{-0.203}^{+0.195}$\\
        \\6& P0304024017& 59276.95& 1528&$0.185_{-0.001}^{+0.001}$& $0.268_{-0.229}^{+0.203}$\\
        \\7& P0304093016& 59278.41& 1395&$0.190_{-0.001}^{+0.001}$& $0.065_{-0.138}^{+0.135}$\\
        \\8& P0304093023& 59292.91& 1380&$0.512_{-0.004}^{+0.006}$& $0.218_{-0.226}^{+0.209}$\\
        \\9& P0304093024& 59295.56& 1967&$0.595_{-0.005}^{+0.007}$& $0.550_{-0.167}^{+0.139}$\\
        \\10& P0304024035& 59295.76& 2084&$0.655_{-0.011}^{+0.012}$& $0.187_{-0.117}^{+0.112}$\\
        \\11& P0304024036& 59297.04& 1200 &$0.902_{-0.013}^{+0.012}$& $0.532_{-0.174}^{+0.145}$\\
        \\12& P0304024038-1&59299.62& 2453&$2.555_{-0.027}^{+0.027}$& $0.257_{-0.186}^{+0.169}$\\
    \hline
    \end{tabular}
\end{table}

\subsection{Energy dependence}

\begin{figure}
\plotone{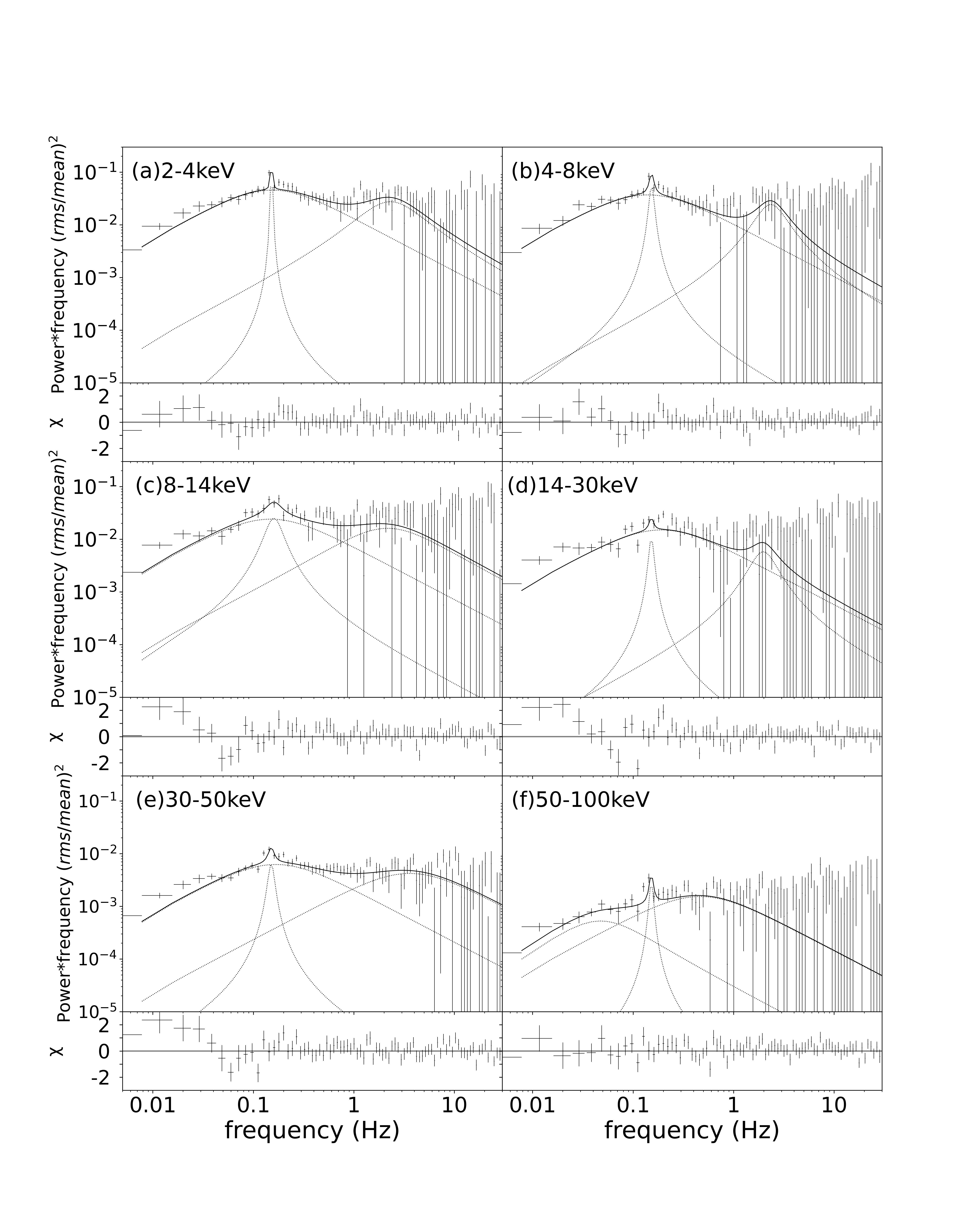}
\caption{The PDS of the observation on MJD 59268 (observation ID P0304024007) in six energy bands: 2-4 keV, 4-8 keV, 8-14 keV, 14-30 kev, 30-50 keV and 50-100 keV.}
\label{fig:pds-energy} 
\end{figure}

\begin{figure}
\plotone{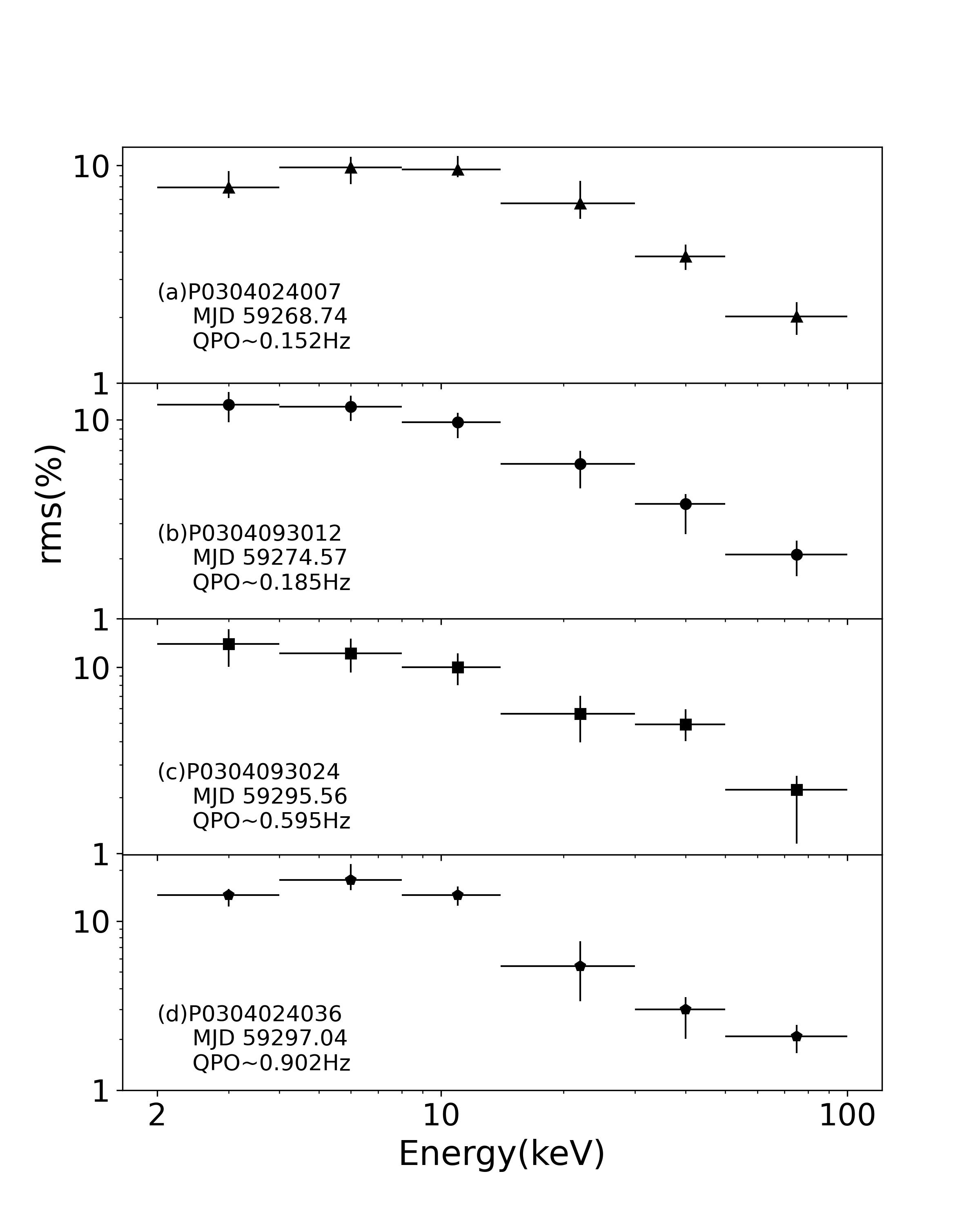}
\caption{The fractional rms amplitude of the type-C QPOs as a function of photon energy of four representative observations from the LHS (a-c) to the transition to the HIMS (d).}
\label{fig:energy}
\end{figure}

In the previous studies of GX $339-4$, the detected QPOs are generally below 50 keV \citep{Nespoli2003,Casella2004,Motta2011,Zhang2017}. To study the evolution of QPOs with energy up to 100 keV, we divide the energy into six bands (i.e., 2--4 keV, 4--8 keV, 8--14 keV, 14--30 keV, 30--50 keV and 50--100 keV) and calculate the properties of QPOs for each new energy band. With the large area of HE detectors, we at the first time detected the LFQPOs in GX 339-4 above 50 keV in several observations. One example is shown in Figure~\ref{fig:pds-energy} (observation ID P0304024007 on MJD 59268), and the fitted results of the same observation are listed in Table~\ref{tab:table2}. We find that as the energy band increases, the QPO centroid frequency is almost unchanged ($\sim$0.152 Hz in the example), while the fractional rms decreases. Figures~\ref{fig:energy} shows the fractional rms of QPOs as a function of photon energy in the four representative observations. The QPO component in the 50-100 keV band cannot be well constrained due to the relatively lower count rate for some of the observations, then we constrain the centroid frequency and the width of QPO in narrow ranges based on the results in the lower energy bands to make the complete rms-energy spectra as shown in Figures~\ref{fig:energy}. The QPO rms remains constant or increases slightly at lower energy bands, but decreases rapidly at higher energies. In general, the QPO rms of GX 339$-$4 decreases from $\sim$ 10\% in 1-2 keV to $\sim$ 2\% in 50-100 keV.

\begin{table}
    \centering
    \caption{LFQPO parameters of different energy bands for the observation on MJD 59268 (ObsID P0304024007).}
    \label{tab:table2}
    \begin{tabular}{ccccc}
    \hline
    \hline
        Telescope&Energy & $\nu_{\rm QPO}$ & rms& Phase lag\\
        &(kev)&(Hz)&(\%)&(rad)\\
    \hline
        \\LE&2-4&$ 0.152\pm0.01$&$ 7.9_{-1.9}^{+1.9}$&reference\\
        \\&4-8&$ 0.153\pm0.01$&$ 9.5_{-2.7}^{+2.1}$&$ -0.094_{-0.144}^{+0.148}$\\
        \hline
        \\ME & 8-14 &$ 0.155\pm0.01$&$ 9.6_{-2.9}^{+2.8}$&$ -0.015_{-0.129}^{+0.128}$\\
        \\&14-30&$ 0.155\pm0.01$&$ 6.7_{-1.4}^{+1.6}$&$ -0.253_{-0.221}^{+0.249}$\\

        \hline
        \\HE & 30-50 &$ 0.150\pm0.01$&$ 3.8_{-0.9}^{+0.8}$&$ -0.182_{-0.123}^{+0.129}$\\
        \\&50-100&$ 0.152\pm0.01$&$ 1.9_{-0.6}^{+0.5}$&$ -0.304_{-0.206}^{+0.236}$\\

    \hline
    \end{tabular}\\
\end{table}

\begin{figure}
\plotone{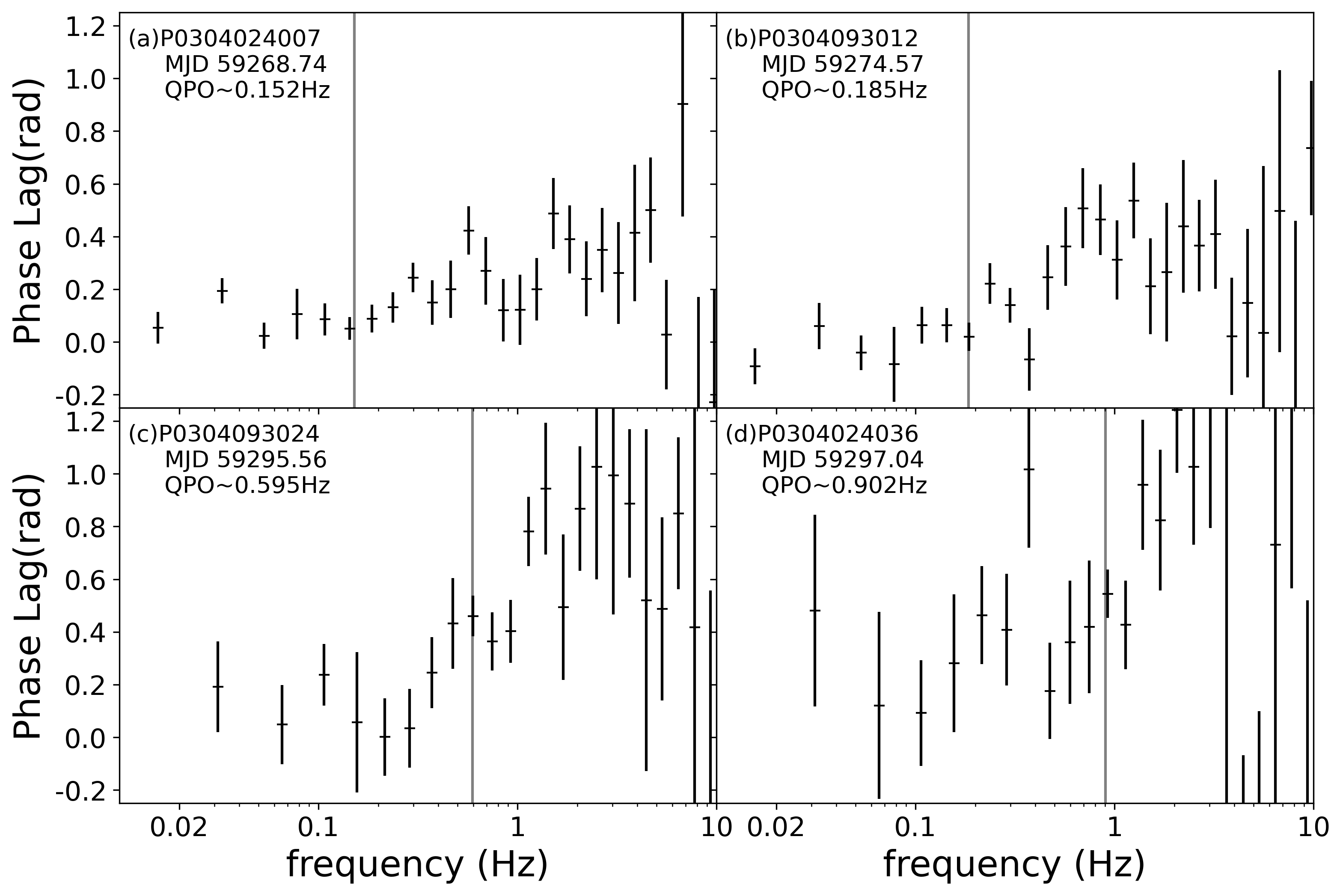}
\caption{The frequency-lag spectra for four representative observations. Phase lags are calculated between 2 - 4 keV and 30 - 50 keV. The grey lines represent the QPO centroid frequencies. The lag here is the combination of the lag of the broad-band noise and the lag of the QPO.}
\label{fig:lag_freq}
\end{figure}

\begin{figure}
\plotone{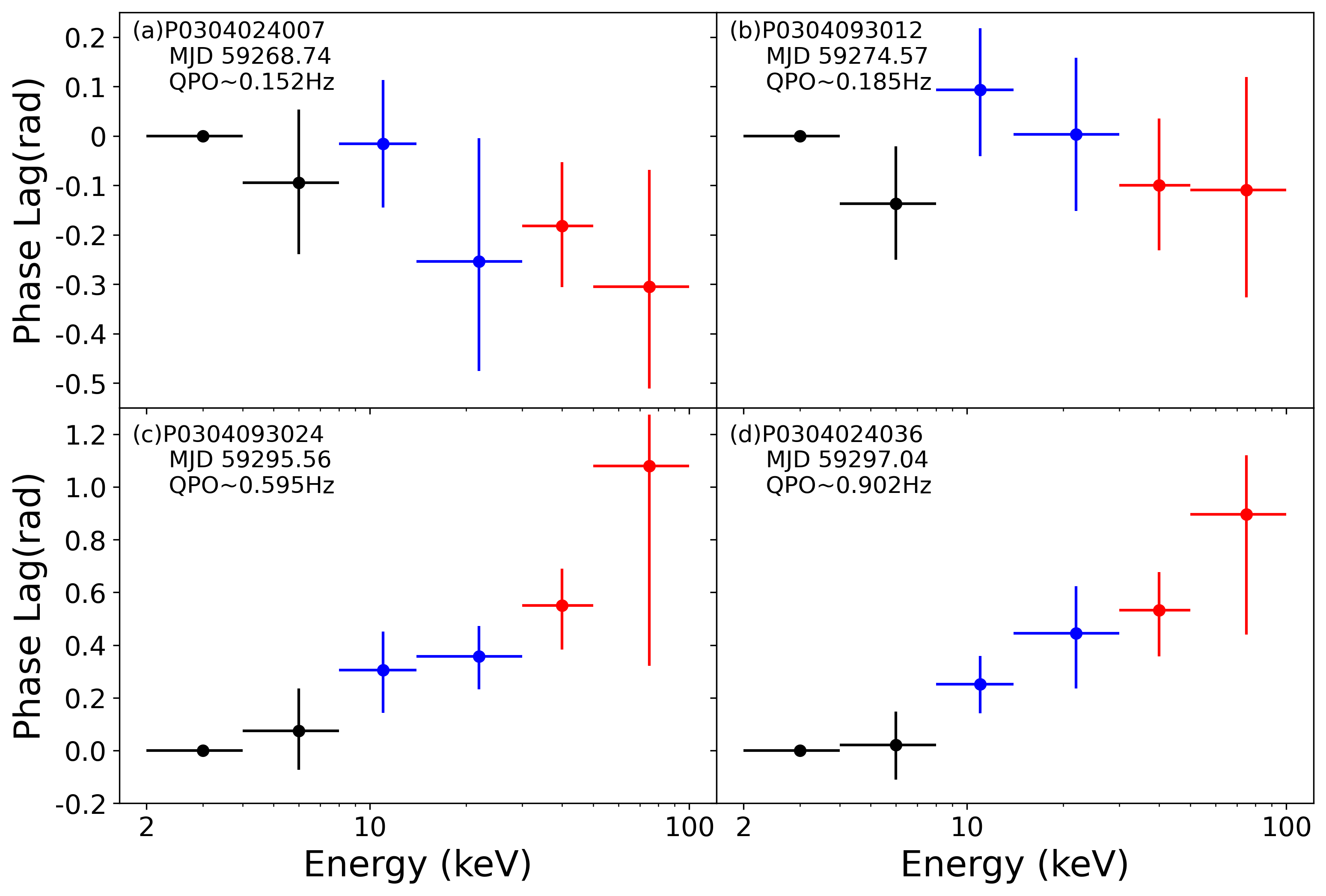}
\caption{LFQPO phase lags of different energy bands for four representative observations.The black, blue, and red points represent the LE, ME, and HE ”intrinsic” lag respectively.}
\label{fig:lag}
\end{figure}

To illustrate the lag dependence around the QPO frequency, we present the frequency-lag spectra between 2 - 4 keV and 30 - 50 keV for four representative observations in Figure~\ref{fig:lag_freq}, where the lag shown here is the combination of the lag of the broad-band noise and the lag of the QPO. In panels (a) and (b) the lag is flat around the QPO frequency, while it shows a increasing trend in panels (c) and (d).
In Figure~\ref{fig:lag}, we then show the energy-dependent QPO lags for the same four observations mentioned above. The black, blue, and red points represent the LE, ME and HE lag respectively. As seen from the plots, the QPO phase lags of panel (a) and (b) during the early stage of the outburst are close to zero with fluctuations from lower to higher energy bands. The lags at higher energy show negative values, but the error bars also become larger. Then the lags gradually become positive as the outburst evolves, especially in the higher energy part above $\sim$ 10 keV, and correlate positively with energy as shown in panel (c) and (d), increasing from $\sim $0.2 rad at $\sim 10$ keV to $\sim $1.0 rad around $50-100$ keV. 

\section{Discussion}
\label{sec:di}

\subsection{Outburst and accretion states}

In this work, we have mainly concentrated on the initial stage of the 2021 outburst in GX 339$-$4, i.e. the LHS and its transition to the HIMS.
During this period, the evolution of count rates shows the totally different trend before and after MJD 59297 (see Figure~\ref{fig:rate}). In addition, after MJD 59297 (from Observation ID P0304024036), the hardness and the intensity no longer wander within a small range but show a rapid change, and the outburst moves to the top left in the HID (Figure~\ref{fig:HID}). More importantly, the observation on MJD 59297 also shows a break-point for the QPO frequency. Before that time, the QPO frequency slightly increases with time without obvious harmonic QPOs, while in the later epoch, rapid changes are seen in the QPO frequencies and harmonic QPOs become significant. The evolution and behavior of the count rates, hardness and the QPO frequency suggest that the transition from LHS to HIMS may begin around MJD 59297. Based on the discussion above, we divide the early 2021 outburst into two stages before and after MJD 59297. In the first stage (before MJD 59297), the source is in the LHS and remains stable for nearly a month. When the transition from LHS to HIMS begins in the second stage (after MJD 59297), the properties of the source (including the count rates, hardness and the QPO frequencies) change significantly within a few days. According to \citet{Liu2022}, the transition may not have been completed until MJD 59299. In addition, \cite{Mondal2023arXiv} provides a further timing study of about one week after MJD 59297 using AstroSat. The frequency of the QPO increases to $\sim$5 Hz, and significant sub-harmonics and harmonics appear.

\cite{Stiele2023} made detailed study of the QPO properties right after our observations between MJD 59300 and MJD 59304 using NICER observations. Their results show that the QPO frequency increases rapidly from $\sim$ 3 Hz to $\sim$ 11 Hz in four days. More intriguingly, they find that the QPO type changes between type-C and type-B for several times, indicating that this outburst undergoes several state transitions between the HIMS and the SIMS. They suggest that the fast transitions are related to the change in the accretion geometry.

A similar evolution was found during the 2006/2007 outburst of GX 339$-$4 \citep{Zhang2017}: the power-law photon index increased slightly and no QPO harmonic was detected in the LHS. When the transition from the LHS to the HIMS started, the power-law photon index increased rapidly accompanied with QPO harmonics and subharmonics appeared.
Comparing the 2021 outburst with the 2006/2007 outburst, a lot of similarities can be found, including the QPO frequency variation and the appearance of harmonic QPOs. This indicates that the evolution processes in the early stage of the two outbursts might be highly similar.

\subsection{Properties of LFQPOs}
We obtain the QPO frequency in the range of $\sim 0.1-3$ Hz (see Table~\ref{tab:table1}). QPOs are normally classified into three types (Type-A, -B, and -C) based on their quality factor Q, the rms, and the shape of the noise associated \citep{Casella2004,Casella2005}. 
The large rms and the Flat-Top Noise presented in the detected QPOs of the 2021 outburst indicate that these QPOs are more likely to be Type-C QPOs \citep{Wijnands1999a,Belloni2002}. Unlike Type-B (during the SIMS) or Type-A (in soft state) QPOs, Type-C QPOs can be detected in all states, but are more prominent in the HIMS \citep{Ingarm2019}.

To explain the origin of LFQPOs (in particular type-C QPOs), different models have been proposed, among which the Lense-Thirring (LT) precession \citep{Stella1998} of the misaligned inner hot flow \citep{Ingram2009,Ingram2010,Ingram2011} is widely accepted, because this model can match the properties of the LFQPOs in BHXRB \citep{Ingram2009}, such as the frequency evolution along the outburst and the rms spectrum \citep{Ingram2011,Ingram2012,Kalamkar2016}. The inward movement of inner flow with increasing mass accretion rates leads to greater illumination and stronger precession of the hot flow \citep{Esin1997,Motta2011}.
In the first stage of the outburst based on our data, the outer radius of inner flow has not decreased yet, causing the QPO frequency and the mass accretion rate to be low. Because of the low X-ray flux and the weak LT precession of the misaligned inner hot flow caused by the low mass accretion rate, an inconspicuous QPO component with low frequency and a strong band-limited noise are shown in the PDS. When the source undergoes the transition from the LHS to the HIMS (the second stage), the mass accretion rate increases and the outer radius of inner flow decreases toward the last stable orbit. Therefore higher QPO frequencies are shown in the PDS (see Figure~\ref{fig:pds-time}).

We have shown the fractional rms of the QPO as a function of photon energy up to 100 keV for four representative observations (see Figure~\ref{fig:energy}). We find that, below 10 keV, the QPO rms amplitude stays stable in the LHS, but decreases monotonically with photon energy beyond 10 keV. The same negative correlation between the rms of QPOs and the photon energy above 10 keV happened during the 2006/2007 outburst when the QPO frequency was less that 1 Hz \citep{Zhang2017}. For other BHXRBs (like GRS 1915+105), the negative correlation was found above 20 keV with RXTE data \citep{Tomsick2001}. 
\cite{Lehr2000} provided a possible explanation of this drop at high energy bands with a Comptonization model that an accretion disk is enshrouded by a corona which is parallel to the plane, assuming that the observed X-ray flux is a combination of the flux from the QPO generation regions and the flux from soft photon unmodulated regions.
The main differences between this model and the LT precession model \citep[i.e.][]{Ingram2009} include that the corona assumed by \cite{Lehr2000} can be outside the inner radius of the truncated disk, which is supported by the current research \citep[e.g.][]{zhang2022MNRAS,García2022}, and the disk temperature gradient assumed does not contradict the LT precession scenario. Based on the model proposed by \cite{Lehr2000}, the QPOs could origin in the disk far from the central object if the high energy drop in amplitude is large.

Previous research shows that GX 339$-$4 has positive lags at the QPO frequenies below 50 keV \citep{Dutta2016,Eijnden2017,Zhang2017,Wang2020}. 
We have calculated the QPO phase lags from 2 - 100 keV in the lag-energy spectra of GX 339$-$4 during its 2021 outburst, as shown in Figure~\ref{fig:lag}. The phase lags in $50-100$ keV provide us more information of this source. According to \cite{Zhang2017}, the lags of GX 339$-$4 exhibit large differences below and above 1.7 Hz. However, we found that observations below 1.7 Hz also show different features. Based on our results below 50 keV, the lags are close to zero with fluctuations when the QPO frequency is $\lesssim$ 0.2 Hz, but rise significantly with increasing energy after 0.2 Hz, which are very similar to the results in \cite{Zhang2017}. 

The positive correlation under 50 keV shown in panels (c) and (d) of Figure~\ref{fig:lag} can be explained by the Comptonization model, as the higher energy photons take a longer path in the corona. The nearly flat lags in {\bf panels} (a) and (b) around zero may be related to the lack of Compton component, as the inner disk radius is large when the QPO frequency is small at the beginning \citep{Ingram2009}, emitting less photons to the corona. In addition, the corona size may vary with QPO frequency \citep{zhang2022MNRAS,zhang2023MNRAS}, and reach its minimum when the lags are approximately flat or zero, which is also confirmed in the source GRS 1915+105 using the spectral-timing comptonization model \citep{García2022,Mendez2022}.

Reflection \citep[e.g.][]{Kotov2001} may also be responsible for the hard lags. Although no reflective component is noticed in our data, it is clearly shown in the lag-energy spectra at high QPO frequencies (\textgreater 1.7 Hz) during the 2006/2007 outburst \citep{Zhang2017}. They believe that the disappearance below 1.7 Hz is either due to its different physical origin, or because the reflection component is weak in the initial stage.

For the first time, we report the 50-100 keV lag behavior in this source. When the QPO centroid frequency is below 0.2 Hz, the lag is negative, but becomes positive after that. Even though this change of sign behavior is not unique \citep[e.g.][]{Eijnden2017,García2022,Mendez2022}, we suggest that this could be due to the large error bars or the imprecise calculation method. The lag after 0.2 Hz is $\sim$ 1.0 rad, which is $\sim$ 180 ms for a 0.9-Hz QPO, or even larger for a lower frequency. The lag is unlikely caused by the propagation of accretion rate fluctuations, since this model predict the time lag on a timescale of seconds \citep{Lyubarskii1997,Uttley2011}, which is larger compared to our results. In addition, it is widely accepted that the QPOs originate from the corona, unless the lag and the QPO are generated in different places. Moreover, the correlation between the photon index and time lags in GX 339-4 is unable to be explained by this model \citep{Kylafis2018AA,Kylafis2020AA}. The Comptonization and reflection models on the other hand, give the hard lags in the order of milliseconds \citep{Zhang2017}, which is much smaller than ours. Again, we can not rule out that the difference is due to large error bars, but even a 0.4 rad lag would cause a time lag of $\sim$ 70 ms. Perhaps the real physical origin of the lags is more complex than we currently imagine, and requires more observational data to verify, especially the data in the high-energy range.

\section{Conclusions}
\label{sec:co}
In this paper, we mainly study the QPOs in the LHS and the transition to the HIMS of GX 339$-$4 during its 2021 outburst. QPOs are detected in most of the observations. The high energy QPOs above 50 keV are discovered at the first time in this source. According to the quality factor and the shape of the noise, we could classify these QPOs as Type-C QPOs. We find that the source stays in the LHS for about a month (MJD 59263 - 59597): the count rate increases slowly with time. During the LHS, the PDS is dominated by a strong band-limited noise and a weak QPO with its centroid frequency increasing over time from $\sim 0.1 - 0.6$ Hz. After MJD 59297, the abrupt change of the count rate and the more distinct QPOs with higher centroid frequency from $0.9-3$ Hz imply that the source is in the transition from the LHS to the HIMS. 

In order to study the energy dependence of QPOs, we compare the properties of QPOs in different energy bands and make the rms-energy and lag-energy spectra. In the rms-energy spectrum, the QPO rms first remains flat at lower energy, and then decreases from $\sim 10\%$ below 10 keV to $2\%$ above 50 keV. 
In the lag-energy spectrum we find the phase lags are first close to zero with some fluctuations, and then become positive and increase with energy up to 100 keV. The large time lags in the band of $50-100$ keV in the later stage of the outburst challenge the existing models. Thus more QPO phase lag studies on high energies are needed. The similarity between our QPO properties and those shown in \cite{Zhang2017} suggests that the 2021 outburst may be highly similar to the 2006/2007 outburst, except that the reflection feature reported in the lag-energy spectra of the earlier outburst is not detected in ours, indicating that the lag properties may change slightly.


\begin{acknowledgments}
\section*{Acknowlgements}
We are grateful to the referee for the useful comments and suggestions to improve the manuscript. This work is supported by the National Key Research and Development Program of China (Grants No. 2021YFA0718500, 2021YFA0718503), the NSFC (12133007, U1838103). This work has made use of data from the \textit{Insight}-HXMT mission, a project funded by China National Space Administration (CNSA) and the Chinese Academy of Sciences (CAS).
\end{acknowledgments}

\bibliography{sample631}{}
\bibliographystyle{aasjournal}



\end{document}